\documentclass[12pt]{article}
\date{}
\begin{document}
\begin{titlepage}
\vspace*{2cm} \baselineskip .5in
\begin{center}
{\Huge\bf A perturbative treatment of a generalized $\mathcal{PT}$
-Symmetric Quartic Anharmonic Oscillator} \vspace{10em}

{\LARGE\bf Abhijit Banerjee\\
\vspace{1em}
\Large\bf Department of Applied Mathematics\\
University of Calcutta\\
Kolkata-700 009, INDIA\\
e-mail : $abhijit_{-}banerjee@hotmail.com$\\} \vspace{6em}
{\Large\bf ABSTRACT\\}
\end{center}
\baselineskip .2in \vspace{2em} \noindent {\large\bf We examine a
generalized $\mathcal{PT}$ -symmetric quartic anharmonic
oscillator model to determine the various physical variables
perturbatively in powers of a small quantity
 $\varepsilon$.We make use of the Bender-Dunne operator basis elements and exploit the properties of the toatally symmetric operator
  $T_{m,n}$.\\\\\\\\
  Journal-Ref. MPLA Vol. 20, No 39 (2005) 3013-3023}
\end{titlepage}
\vspace{1em} \setcounter{page}{2} \baselineskip .3in
 \noindent{\large\bf  1.\quad
Introduction:} \\During recent years $\mathcal{PT}$ -symmetric
quantum mechanics has emerged as an area of high theoretical
interest (e.g.,[1-10] and references therein). For one thing,
$\mathcal{PT}$ -symmetry is a  weaker condition compared to the
usual Hermiticity but exhibits all the essential properties of a
Hermitian quantum Hamiltonian. For another, $\mathcal{PT}$
-symmetry opens up the window to the non-Hermitian world, thus
enabling one to address a much broader class of
Hamiltonians.\\\indent
 Although the current interest in $\mathcal{PT}$ -symmetry stems from the 1998 seminal paper of
Bender and Boettcher [1] where it was shown that for a certain
class of $\mathcal{PT}$ -symmetric Hamiltonians the spectrum
remained entirely real, discrete and bounded below, the concept of
$\mathcal{PT}$ -symmetry had its roots in some earlier independent
works as well. These include the ones of Caliceti et al [11] and
Bessis and Zinn-Justin  who studied a cubic anharmonic oscillator
model with an imaginery coupling and that of Buslaev and Greechi
[12] who analysed the spectra of certain non-Hermitian versions of
the quartic anharmonic oscillator.\\\indent
 Recently Mostafazadeh [13] has revisited the question of observables
 for the
$\mathcal{PT}$ -symmetric cubic anharmonic oscillator problem and,
in this regard, has performed a perturbative calculation of the
physical observables including investigation of the classical
limit. Motivated by Mostafazadeh's work, we examine,in this note,
the $\mathcal{PT}$ -symmetric version of a generalized quartic
anharmonic oscillator described by the Hamiltonian
$$H=\frac{p^{2}}{2m}+\frac{\mu^{2}}{2}x^{2}+i\epsilon
x^{3}-m\hbar^{2}\epsilon^{2}x^{4}\eqno(1)$$with  $(\mu,\nu \in R)$
that includes a cubic anharmonicity as well.
\\\indent Noting that a $\mathcal{C}$ -operator can be introduced
[14] in the physical Hilbert space $\mathcal{H}_{phys}$ subject to
a $\mathcal{CPT}$ -inner product [15], and that it commutes with
both H and
 $\mathcal{PT}$ , we show that for the above H an equivalent
 Hermitian Hamiltonian h can be set up. The classical
 Hamiltonian $H_{c}$ is then obtained in the limit $\hbar \rightarrow
 0$.The physical position and momentum operators $\large\bf X$ and
  $\large\bf P$ ,which are  actually $\eta_{+}$ -pseudo-Hermitian for the
  metric operator $\eta_{+}$ and related to the conventional
  position (x) and momentum (p) operators by the same similarity
  transformation that links H and h,clearly turns out to be  $\mathcal{PT}$
  -symmetric,a result similar to the  $\mathcal{PT}$ -symmetric
  cubic oscillator. We also calculate the eigenvalues of H based
on the first-order Rayleigh-Schr\"{o}dinger perturbation theory
upto and including terms of order $\epsilon^{3}$.
  Further we determine the conserved probability
  density for a given state vector $\psi\in\mathcal{H}_{phys}$.
  It should be mentioned that our calculations are somewhat
  different from Mostafazadeh's in that we have exploited the
  symmetrized objects $T_{m,n}$ [17] satisfying commutation
  (lowering type) and anti-commutation (raising type) relations to
  write down the perturbative expansion of the $\mathcal{C}$
  -operator.\\\\\\
\noindent{\large\bf  2.\quad Basic equations:}

\hspace{1em}To ensure the reality of the spectrum of a
diagonalizable operator it is necessary that the Hamiltonian $H$
must be Hermitian with respect to a positive definite inner
product $<.,.>_{+}$.
 The latter can be expressed in terms of a positive definite metric
operator $\eta_{+}:\mathcal{H}\rightarrow\mathcal{H}$ of the
reference Hilbert space $\mathcal{H}$ in which $H$
acts:$$<.,.>_{+}=<.,\eta_{+}.>\eqno(2)$$ where $\eta_{+}$ belongs
to the set of all Hermitian invertiable operators
 $\eta:\mathcal{H}\rightarrow\mathcal{H}$ satisfying
 $H^{\dag}=\eta H \eta^{-1}$ [18] and can be expressed as
$$\eta_{+}=e^{-Q}\eqno(3)$$ where Q is Hermitian. In terms of
$\eta_{+}$, $\mathcal{C}$ admits a representation
$$\mathcal{C} = \mathcal{P}\eta_{+} =
\eta_{+}^{-1}\mathcal{P}\eqno(4)$$ The operator $\mathcal{C}$
commutes with both H and $\mathcal{PT}$ : $[\mathcal{C},H]=0$
,$[\mathcal{C},\mathcal{PT}]=0$ and mimicks the charge conjugation
operator in particle theory.
\\\indent Any
Hermitian physical observable $\mathcal{O}\in\mathcal{H}_{phys}$
can be converted to a Hermitian operator $o\in\mathcal{H}$ by the
transformation
$$\mathcal{O}=\rho^{-1}o\rho\eqno(5)$$
where $\rho=\surd{\eta_{+}}$ is a unitary operator and because of
(3) may be given by
$$\rho=e^{-Q/2}\eqno(6)$$\\\indent In view of (5) we can write
 $$H=\rho^{-1} h \rho\eqno(7)$$ where h is the corresponding
Hermitian Hamiltonian. The classical Hamiltonian  $H_{c}(\bf
x_{c},\bf p_{c})$ is obtained
 from $h({\bf x_{c}},{\bf p_{c}})$ by the relation
 $$H_{c}({\bf x_{c}},{\bf p_{c}})=\lim_{\hbar \rightarrow 0}
 h(\bf x_{c},\bf p_{c})\eqno(8)$$
 where the limit is assumed to exist.\\\indent
For the sake of convenience let us introduce a set of new
variables
$$ X: = \hbar^{-1}{\bf x},\hspace{1em}
 P:= {\bf p},\hspace{1em}
\mathcal{M}:= m^{1/2}\hbar\mu,\hspace{1em}
\varepsilon:=m\hbar^{3}\epsilon \eqno(9)$$ In terms of X and P,
$H({\bf x},{\bf p})\rightarrow H(X,P)$ with $$H(X,P)=
H_{0}(X,P)+\varepsilon H_{1}(X,P)+\varepsilon^{2} H_{2}(X,P)\eqno(10)$$\\
$$\begin{array}{lllll}& &
H_{0}(X,P)=\frac{1}{2}P^{2}+\frac{1}{2}\mathcal{M}^{2}X^{2}
&(11)&\\\\
& & H_{1}(X,P)= iX^{3} &(12)&\\\\
& & H_{2}(X,P)= -X^{4} &(13)&\\\\
& & H(X,P)= m H(\bf x,\bf p) &(14)&\\\\
\end{array}$$\\\\
 \noindent{\large\bf  3.\quad Determining
 the Q and  $\mathcal{C}$-operators:} \\\indent Using (3)in (4), we consider the general form
of
 $\mathcal{C}$ as
 $$\mathcal{C}=e^{Q(X,P)}\mathcal{P}\eqno(15)$$
 It has the following properties:
 $$[\mathcal{C},\mathcal{P}\mathcal{T}]= 0\eqno(16)$$
 $$\mathcal{C}^{2}= 1\eqno(17)$$
 $$[\mathcal{C},H(X,P)]= 0\eqno(18)$$
but $[\mathcal{C},\mathcal{P}]\neq 0$ and
  $[\mathcal{C},\mathcal{T}]\neq 0$\\\indent
 Substitution of $\mathcal{C}$ from (15) into
 (16)implies
 $$e^{Q(X,P)}\mathcal{P}\mathcal{P}\mathcal{T}=\mathcal{P}\mathcal{T}e^{Q(X,P)}\mathcal{P}\eqno(19)$$
 showing
 $Q(X,P)$ to be an even function of X: $ e^{Q(X,P)}=e^{Q(-X,P)}$
 .That $Q(X,P)$ is an odd function of P follows from the
 consideration (17):
 $$e^{Q(X,P)}\mathcal{P}e^{Q(X,P)}\mathcal{P}=1\eqno(20)$$
 which yields $e^{Q(X,P)}=e^{-Q(-X,-P)}$ .\\\indent
 We now expand of $Q(X,P)$ in a series of
odd powers of  $\varepsilon$ ,  namely $$Q(X,P)=\varepsilon
Q_{1}(X,P)+\varepsilon ^{3}Q_{3}(X,P)+\varepsilon
^{5}Q_{5}(X,P)+\varepsilon ^{7}Q_{7}(X,P)+ O(\varepsilon
^{8})\eqno(21)$$ Using(18),it follows that
$$e^{Q(X,P)}H_{0}-H_{0}e^{Q(X,P)}=\varepsilon(e^{Q(X,P)}H_{1}+H_{1}e^{Q(X,P)})-\varepsilon^{2}(e^{Q(X,P)}H_{2}-H_{2}e^{Q(X,P)})$$
Left multiplying both sides leads to,
$$H_{0}-e^{-Q(X,P)}H_{0}e^{Q(X,P)}=\varepsilon(H_{1}+
e^{-Q(X,P)}H_{1}e^{Q(X,P)})-\varepsilon^{2}(H_{2}-e^{-Q(X,P)}H_{2}e^{Q(X,P)})\eqno(22)$$
\\\indent Using Baker-Campbell-Hausdorff identity, i.e,
$$e^{-A}Be^{A}=B+[B,A]+\frac{1}{2!}[[B,A],A]+\frac{1}{3!}[[[B,A],A],A]+....\eqno(23)$$ we can arrange the
above expression as
$$\begin{array} {lll}&
&-[H_{0},Q]-\frac{1}{2!}[[H_{0},Q],Q]-\frac{1}{3!}[[[H_{0},Q],Q],Q]-\frac{1}{4!}[[[[H_{0},Q],Q],Q],Q]\\\\&
&-\frac{1}{5!}[[[[[H_{0},Q],Q],Q],Q],Q]

-\frac{1}{6!}[[[[[[H_{0},Q],Q],Q],Q],Q],Q]\\\\& &-\frac{1}{7!}[[[[[[[H_{0},Q],Q],Q],Q],Q],Q],Q]\\\\
&=& 2\varepsilon
H_{1}+\varepsilon[H_{1},Q]+\frac{\varepsilon}{2!}[[H_{1},Q],Q]+\frac{\varepsilon}{3!}[[[H_{1},Q],Q],Q]+\frac{\varepsilon}{4!}[[[[H_{1},Q],Q],Q],Q]\\\\
&
&+\frac{\varepsilon}{5!}[[[[[H_{1},Q],Q],Q],Q],Q]+\frac{\varepsilon}{6!}[[[[[[H_{1},Q],Q],Q],Q],Q],Q]\\\\&
&+\frac{\varepsilon}{7!}[[[[[[[H_{1},Q],Q],Q],Q],Q],Q],Q]
+\varepsilon^{2}[H_{2},Q]+\frac{\varepsilon^{2}}{2!}[[H_{2},Q],Q]+\frac{\varepsilon^{2}}{3!}[[[H_{2},Q],Q],Q]\\\\&
&+\frac{\varepsilon^{2}}{4!}[[[[H_{2},Q],Q],Q],Q]
+\frac{\varepsilon^{2}}{5!}[[[[[H_{2},Q],Q],Q],Q],Q]\\\\&
&+\frac{\varepsilon^{2}}{6!}[[[[[[H_{2},Q],Q],Q],Q],Q],Q]
+\frac{\varepsilon^{2}}{7!}[[[[[[[H_{2},Q],Q],Q],Q],Q],Q],Q]\hspace{2em}(24)\end{array}$$
where we have taken the terms upto order
$\varepsilon^{7}$\\\indent Substituting (21) into (24) and
equating terms of order
 $\varepsilon,\varepsilon^{3},\varepsilon^{5},\varepsilon^{7}$ we
get,
$$\begin{array}{lllll}
[H_{0},Q_{1}]&=& -2H_{1} &(25)&\\\\

[H_{0},Q_{3}]&=& -\frac{1}{6}[Q_{1},[Q_{1},H_{1}]]+[Q_{1},H_{2}]&(26)&\\\\

[H_{0},Q_{5}]&=&-\frac{1}{6}([Q_{3},[Q_{1},H_{1}]]+[Q_{1},[Q_{3},H_{1}]])+\frac{1}{360}[Q_{1},[Q_{1},[Q_{1},[Q_{1},H_{1}]]]]+[Q_{3},H_{2}]&(27)&\\\\

[H_{0},Q_{7}]&=& -\frac{1}{6}([Q_{5},[Q_{1},H_{1}]]+[Q_{3},[Q_{3},H_{1}]]+[Q_{1},[Q_{5},H_{1}]])\\\\
& &+\frac{1}{360}([Q_{3},[Q_{1},[Q_{1},[Q_{1},H_{1}]]]]+[Q_{1},[Q_{3},[Q_{1},[Q_{1},H_{1}]]]]\\\\

& &+[Q_{1},[Q_{1},[Q_{3},[Q_{1},H_{1}]]]]+[Q_{1},[Q_{1},[Q_{1},[Q_{3},H_{1}]]]])\\\\
&
&-\frac{1}{15120}[Q_{1},[Q_{1},[Q_{1},[Q_{1},[Q_{1},[Q_{1},H_{1}]]]]]]+[Q_{5},H_{2}]&(28)&\end{array}$$
note that terms of order
 $\varepsilon^{2},\varepsilon^{4},\varepsilon^{6},$ i.e, even
 powers of $\varepsilon$ give no new results.\\\indent
 To solve for (25),(26),(27)and (28) we introduce, following Bender and Dunne [16] the totally symmetrized sum  $T_{r,s}$ over all
 terms containing r-factor of P and s-factor of X.
 For example,we have $$\begin{array}{lll}T_{0,0}&=&1\\\\
 T_{1,0}&=&P\\\\T_{1,2}&=&\frac{1}{3}(PX^{2}+XPX+X^{2}P)\\\\
  T_{0,3}&=&X^{3}\\\\ T_{3,1}&=&\frac{1}{4}(XP^{3}+PXP^{2}+P^{2}XP+P^{3}X)
 \\\\T_{2,2}&=&\frac{1}{6}(X^{2}P^{2}+P^{2}X^{2}+PXPX+XPXP+PX^{2}P+XP^{2}X)\end{array}$$and
 so on.\\\\\indent We thus get
 $$\begin{array}{lllll}Q_{1}&=&-\frac{4}{3}\mathcal{M}^{-4}T_{3,0}-2\mathcal{M}^{-2}T_{1,2} &(28a)&\\\\
Q_{3}&=&(\frac{128}{15}\mathcal{M}^{-10}-\frac{32}{5}\mathcal{M}^{-8})T_{5,0}+(\frac{40}{3}\mathcal{M}^{-8}\\\\& &-16\mathcal{M}^{-6})T_{3,2}+(8\mathcal{M}^{-6}-8\mathcal{M}^{-4})T_{1,4}-(12\mathcal{M}^{-8}-8\mathcal{M}^{-6})T_{1,0}&(28b)&\\\\
Q_{5}&=&(\frac{6368}{15}\mathcal{M}^{-12}-128\mathcal{M}^{-10}+128\mathcal{M}^{-8})T_{1,2}+(-64\mathcal{M}^{-10}+32\mathcal{M}^{-8}-32\mathcal{M}^{-6})T_{1,6}\\\\
& &+(\frac{24736}{45}\mathcal{M}^{-14}-256\mathcal{M}^{-12}+\frac{640}{3}\mathcal{M}^{-10})T_{3,0}+(-\frac{512}{3}\mathcal{M}^{-12}+\frac{352}{3}\mathcal{M}^{-10}-128\mathcal{M}^{-8})T_{3,4}\\\\
& &+(-\frac{544}{3}\mathcal{M}^{-14}+128\mathcal{M}^{-12}-128\mathcal{M}^{-10})T_{5,2}+(-\frac{320}{3}\mathcal{M}^{-16}+\frac{256}{7}\mathcal{M}^{-14}-\frac{256}{7}\mathcal{M}^{-12})T_{7,0}&(28c)&\\\\
Q_{7}&=&(\frac{553984}{315}\mathcal{M}^{-22}-\frac{124416}{315}\mathcal{M}^{-20}+\frac{69632}{315}\mathcal{M}^{-18}-\frac{2048}{9}\mathcal{M}^{-16})T_{9,0}\\\\
& &+(\frac{97792}{35}\mathcal{M}^{-20}-\frac{62208}{35}\mathcal{M}^{-18}+\frac{34816}{35}\mathcal{M}^{-16}-1024\mathcal{M}^{-14})T_{7,2}\\\\
&
&+(\frac{377344}{105}\mathcal{M}^{-18}-\frac{35456}{15}\mathcal{M}^{-16}+\frac{7424}{5}\mathcal{M}^{-14}-1536\mathcal{M}^{-12})T_{5,4}\\\\
&
&+(\frac{721024}{315}\mathcal{M}^{-16}-\frac{4096}{3}\mathcal{M}^{-14}+\frac{2432}{3}\mathcal{M}^{-12}-\frac{2560}{3}\mathcal{M}^{-10})T_{3,6}\\\\
&
&+(\frac{1792}{3}\mathcal{M}^{-14}-256\mathcal{M}^{-12}+128\mathcal{M}^{-10}-128\mathcal{M}^{-8})T_{1,8}\\\\
&
&+(-\frac{2209024}{105}\mathcal{M}^{-20}+\frac{619648}{75}\mathcal{M}^{-18}-\frac{54272}{15}\mathcal{M}^{-16}+3584\mathcal{M}^{-14})T_{5,0}$$\\\\
&
&+(-\frac{2875648}{105}\mathcal{M}^{-18}+\frac{141824}{15}\mathcal{M}^{-16}-\frac{15616}{3}\mathcal{M}^{-14}+5120\mathcal{M}^{-12})T_{3,2}\\\\
&
&+(-\frac{390336}{35}\mathcal{M}^{-16}+\frac{40832}{15}\mathcal{M}^{-14}-1216\mathcal{M}^{-12}+1280\mathcal{M}^{-10})T_{1,4}\\\\
&
&+(\frac{46976}{5}\mathcal{M}^{-18}-\frac{49472}{15}\mathcal{M}^{-16}+1536\mathcal{M}^{-14}-1280\mathcal{M}^{-12})T_{1,0}&(28d)&\end{array}$$
Explicit forms of $Q(X,P)$ and $\mathcal{C}(X,P)$ are then
obtained by substituting the above expressions for
$Q_{1},Q_{3},Q_{5},Q_{7}$ into (21) and (15).\\\\
\noindent{\large\bf  4.\quad Determining the $\large\bf X$ and
$\large\bf P$ operators:}\\\indent
 Now we calculate the physical
position and momentum operator
 $\large\bf X$ and $\large\bf P$ from the previously introduced variables
 X,P using the similarity transformation (5):
 $${\large\bf X}=\rho^{-1}X\rho
 =e^{\frac{Q}{2}}Xe^{-\frac{Q}{2}}\eqno(29)$$
$${\large\bf P}=\rho^{-1}P\rho
 =e^{\frac{Q}{2}}Pe^{-\frac{Q}{2}}\eqno(30)$$\\\indent
 Using (23),we obtain for $\large\bf X$ and
 $\large\bf P$
$$\begin{array}{lll}{\large\bf
X}&=&X+\varepsilon(2i\mathcal{M}^{-4}P^{2}+i\mathcal{M}^{-2}X^{2})+\varepsilon^{2}(2\mathcal{M}^{-6}XP^{2}-2i\mathcal{M}^{-6}P-\mathcal{M}^{-4}X^{3})\\\\
&
&+\varepsilon^{3}[(-\frac{172}{15}\mathcal{M}^{-10}+16\mathcal{M}^{-8})iP^{4}-(5\mathcal{M}^{-6}-4\mathcal{M}^{-4})iX^{4}\\\\&
&-(\frac{128}{3}\mathcal{M}^{-8}-48\mathcal{M}^{-6})XP
+(\frac{64}{3}\mathcal{M}^{-8}-24\mathcal{M}^{-6})iX^{2}P^{2}\\\\&
&+(\frac{50}{3}\mathcal{M}^{-8}-16\mathcal{M}^{-6})i]+O(\varepsilon^{4})\hspace{2em}(31)\\\\
{\large\bf
P}&=&P-\varepsilon(2i\mathcal{M}^{-2}(XP-\frac{i}{2}))+\varepsilon^{2}(2\mathcal{M}^{-6}P^{3}-\mathcal{M}^{-4}(X^{2}P-iX))\\\\
&
&-i\varepsilon^{3}[(16\mathcal{M}^{-8}-16\mathcal{M}^{-6})(XP^{3}-\frac{3}{2}iP^{2})\\\\&
&+(16\mathcal{M}^{-6}-16\mathcal{M}^{-4})(X^{3}P-\frac{3}{2}iX^{2})]+O(\varepsilon^{4})\hspace{2em}(32)\end{array}$$\\\indent
where we retained terms of order of $\varepsilon^{3}$.\\From (31)
and (32),we easily see that
$$\mathcal{P}{\large\bf X}\mathcal{P}\neq{-\large\bf X}$$
$$\mathcal{P}{\large\bf P}\mathcal{P}\neq{-\large\bf P}$$
$$\mathcal{T}{\large\bf X}\mathcal{T}\neq{\large\bf X}$$
$$\mathcal{T}{\large\bf P}\mathcal{T}\neq{-\large\bf P}$$
but
$$\mathcal{PT}{\large\bf X}\mathcal{PT}={-\large\bf
X}$$
$$\mathcal{PT}{\large\bf P}\mathcal{PT}={\large\bf
P}\eqno(33)$$ From (33) we conclude that the physical position and
momentum operator,i.e, {\large\bf X} and {\large\bf P} $\in$
$\mathcal{H}_{phys}$ are $\mathcal{PT}$-symmetric.\\\\ `
\noindent{\large\bf  5.\quad The equivalent Hermitian
Hamiltonian:}\\\indent
 For the operator $\rho$,
the corresponding Hermitian Hamiltonian ${\bf h}(X,P)$ is given
according to (7)with previously introduced variables X,P as$${\bf
h}(X,P)=e^{-Q/2} {H(X,P)} e^{Q/2}\eqno(34)$$\\\indent Introducing
the perturbative expansion for ${\bf h}(X,P)$ as
$${\bf h}=\sum^{\infty}_{i=0}h^{(i)}\varepsilon^{i}\eqno(35)$$
and using (10),(21),(25)-(28) and (34) along with the
Baker-Campbell-Hausdorff identity(23), we obtain for the various
coefficients
 $h^{(i)}$ ,i=0,1,2,3,...., the results
$$\begin{array}{lll}h^{(0)}&=&H_{0}\\\\
h^{(2)}&=&H_{2}+\frac{1}{4}[H_{1},Q_{1}]\\\\
h^{(4)}&=&\frac{1}{4}[H_{1},Q_{3}]-\frac{1}{192}[[[H_{1},Q_{1}],Q_{1}],Q_{1}]\\\\
h^{(6)}&=&\frac{1}{4}[H_{1},Q_{5}]-\frac{1}{192}([[[H_{1},Q_{1}],Q_{1}],Q_{3}]+[[[H_{1},Q_{1}],Q_{3}],Q_{1}]\\\\&
&+[[[H_{1},Q_{3}],Q_{1}],Q_{1}])+\frac{1}{7680}[[[[[H_{1},Q_{1}],Q_{1}],Q_{1}],Q_{1}],Q_{1}]\hspace{5em}(36)\end{array}$$
with the odd ones vanishing: $h^{(i)}=0$ ,i=1,3,5,7,....\\
\\\indent
Keeping terms up to the order $\varepsilon^{5}$ ; ${{\bf h}(X,P)}$
can thus be expressed as\\
$$\begin{array}{lll}{{\bf
h}(X,P)}&=&\frac{1}{2}(T_{2,0}+\mathcal{M}^{2}T_{0,2})+\varepsilon^{2}[-\frac{1}{2}\mathcal{M}^{-4}+(\frac{3}{2}\mathcal{M}^{-2}-1)T_{0,4}+3\mathcal{M}^{-4}T_{2,2}]\\\\
&
&+\varepsilon^{4}[-(36\mathcal{M}^{-10}-24\mathcal{M}^{-8})T_{4,2}+(27\mathcal{M}^{-10}-24\mathcal{M}^{-8})T_{2,0}-(\frac{51}{2}\mathcal{M}^{-8}-36\mathcal{M}^{-6})T_{2,4}\\\\
&
&+(\frac{179}{24}\mathcal{M}^{-8}-12\mathcal{M}^{-6})T_{0,2}-(\frac{7}{2}\mathcal{M}^{-6}-6\mathcal{M}^{-4})T_{0,6}+2\mathcal{M}^{-12}T_{6,0}]+O(\varepsilon^{6})\hspace{4em}(37)\end{array}$$
\\If now we consider the normalized eigen vector $|n>$ of the
conventional Harmonic oscillator $H_{0}$, then we can easily
calculate $E_{n}$ for $\bf H$ by the first order
Rayleigh-Schr\"{o}dinger perturbation theory. We obtain upto the
terms of order $\varepsilon^{3}$\\
$$\begin{array}{lll}E_{n}&=&\mathcal{M}(n+\frac{1}{2})+\varepsilon^{2}<n|h^{(2)}|n>+O(\varepsilon^{4})\\\\
&
&=\mathcal{M}(n+\frac{1}{2})+\frac{\varepsilon^{2}}{4}[\frac{1}{2\mathcal{M}^{4}}(30n^{2}+30n+11)-(6n^{2}+6n+3)]+O(\varepsilon^{4})\hspace{4em}(38)\end{array}$$
\\\\\noindent{\large\bf  6.\quad Classical Hamiltonian:}\\\indent
Employing (9) , we have,$$T_{r,s}=\hbar^{-s}S_{r,s}\eqno(39)$$
where $S_{r,s}$ be the totally symmetrized sum of all terms
containing r-factor of p and s-factor of x. Specifically
$$\begin{array}{lll}S_{0,0}&=&1\\\\
S_{0,1}&=&x\\\\
S_{3,0}&=&p^{3}\\\\
S_{1,3}&=&\frac{1}{4}(x^{3}p+xpx^{2}+x^{2}px+px^{3})\end{array}$$
and so on.\\\indent Now from (14),$${\bf
h}(X,P)=mh(x,p)\eqno(40)$$ Keeping terms up to the terms of order
of $\varepsilon^{3}$ in (37) and using the relations (9),(39),(40)
we finally obtain
$$h(x,p)=\frac{p^{2}}{2m}+\frac{1}{2}\mu^{2}x^{2}+\epsilon^{2}m[(\frac{3}{2}m^{-1}\mu^{-2}-\hbar^{2})x^{4}
-2m^{-2}\hbar^{2}\mu^{-4}-6im^{-2}\hbar\mu^{-4}xp-3m^{-2}\mu^{-4}x^{2}p^{2}]+O(\epsilon^{4})\eqno(41)$$\\\indent
The corresponding classical Hamiltonian can be read off from (8):
$$\begin{array}{lll}H_{c}(x_{c},p_{c})&=&\lim_{\hbar \rightarrow
 0}h(x_{c},p_{c})\\\\
 &=&\frac{p_{c}^{2}}{2M(x_{c})}+\frac{1}{2}\mu^{2}x_{c}^{2}+\frac{3\epsilon^{2}}{2\mu^{2}}x_{c}^{4}+O(\epsilon^{4})\hspace{4em}(42)\end{array}$$
 where
 $$M(x_{c})=\frac{m}{1-6\mu^{-4}\epsilon^{2}x_{c}^{2}}\eqno(43)$$\\
A position-dependent mass $M(x_{c})$ is implied by (43) for the
classical particle whose dynamics is dictated by the Hamiltonian
 $H_{c}(x_{c},p_{c})$.\\\\
\noindent{\large\bf  7.\quad Conserved probability
density:}\\\indent For a given state vector
$\psi\in\mathcal{H}_{phys}$, the perturbation expansion for the
corresponding physical wave function is
$$\begin{array}{lll}\Psi(x)&=&<x\mid e^{-\frac{Q}{2}}\mid \psi>\\\\
&=&<x\mid \sum^{\infty}_{k=0}\frac{(-1)^{k}Q^{k}}{2^{k}k!}\mid
\psi>\\\\
&=&\psi(x)+<x\mid -\frac{Q_{1}}{2}\mid \psi>\varepsilon+<x\mid
\frac{Q_{1}^{2}}{8}\mid \psi>\varepsilon^{2}+<x\mid
(-\frac{Q_{3}}{2}-\frac{Q_{1}^{3}}{48}) \mid
\psi>\varepsilon^{3}+O(\varepsilon^{4})\hspace{2em}(44)\end{array}$$\\\indent
Using (9),(28a) and (28b) we obtain from (44)
$$\Psi(x)=(1+\epsilon L_{1}+\epsilon ^{2}L_{2}+\epsilon^{3}
L_{3})\psi(x)+O(\epsilon^{4})\eqno(45)$$ where
$$\begin{array}{lllll}L_{1}&=&-\frac{m}{2}\hbar^{3}\hat{Q_{1}} &(46)&\\\\
L_{2}&=&\frac{m^{2}}{8}\hbar^{6}\hat{Q_{1}}^{2}&(47)&\\\\
L_{3}&=&-\frac{m^{3}}{2}\hbar^{9}\hat{Q_{3}}-\frac{m^{3}}{48}\hbar^{9}\hat{Q_{1}}^{3}&(48)&\end{array}$$
and,
$$\begin{array}{lllll}\hat{Q_{1}}&=&-\frac{4}{3}m^{-2}\hbar^{-4}\mu^{-4}S_{3,0}-2m^{-1}\hbar^{-2}\mu^{-2}S_{1,2}&(49)&\\\\
\hat{Q_{3}}&=&(\frac{128}{15}m^{-5}\hbar^{-10}\mu^{-10}-\frac{32}{5}m^{-4}\hbar^{-8}\mu^{-8})S_{5,0}+(\frac{40}{3}m^{-4}\hbar^{-10}\mu^{-8}-16m^{-3}\hbar^{-8}\mu^{-6})S_{3,2}\\\\
&
&+(8m^{-3}\hbar^{-10}\mu^{-6}-8m^{-2}\hbar^{-8}\mu^{-4})S_{1,4}-(12m^{-4}\hbar^{-8}\mu^{-8}-8m^{-3}\hbar^{-6}\mu^{-6})S_{1,0}&(50)&\end{array}$$\\\indent
Employing (46)-(50) into (45), we find the conserved probability
density $\varrho$ associated with a given state vector
$\psi\in\mathcal{H}_{phys}$ as
$$\varrho(x)=N^{-1}\mid\Psi(x)\mid^{2}\eqno(51)$$
where $$N=\int^{\infty}_{-\infty}\mid\Psi(x)\mid^{2}dx\eqno(52)$$
\\\noindent{\large\bf  8.\quad Conclusion:}\\\indent
We have carried out a perturbative treatment to study a
$\mathcal{PT}$ -symmetric quartic anharmonic oscillator model. We
have shown possible to set up an equivalent Hermitian Hamiltonian
by employing a similarity transformation. Such a Hamiltonian has a
classical limit too. Physical position and momentum operators have
been determined perturbatively and energy eigenvalues are obtained
in the framework of first order Rayleigh-Schr\"{o}dinger
perturbation theory. In all these calculations we have kept terms
up to and including those of order $\varepsilon^{3}$.  The
conserved probability density is also determined. Finally,let us
mention that in the absence of the quartic term in (1) our results
essentially reduce to those of Mostafazadeh's [13] for the
physical observables ${\large\bf X}$ and ${\large\bf P}$,
equivalent Hermitian Hamiltonian $h(X,P)$ and the energy spectrum
derived from the first order Rayleigh-Schr\"{o}dinger perturbation
theory.\\\\

\noindent {\large\bf Acknowledgement}\\\indent I would like to
thank Prof. Bijan Kumar Bagchi, Department of Applied Mathematics,
University of Calcutta for his continuous help and proper
guidance.I would also like to thank Prof. A.Mostafazadeh,Ko\c{c}
University,for a helpful correspondence.
 I am grateful to the
  University Grant Commission, New Delhi, India for
financial support.\\\\
\noindent {\large\bf References:}

\vspace{1em}

\begin{enumerate}
\item Bender C M and Boettcher S
\hspace{1em}1998 Phys.Rev.Lett. {\bf24} 5243\\
See also\\ Bender C M\hspace{1em}-Introduction to
$\mathcal{PT}$-symmetric quantum theory,preprint quant-ph/0501052
\item Delabaere E and Pham F\hspace{1em}1998 Phys.Lett.A {\bf250} 25\\
Fernandez F M, Guardiola R, Ros J and Znojil M\hspace{1em}1998
J.Phys.A: Math.Gen. {\bf31} 10105 \\
Bender C M, Boettcher S and Meisinger P N\hspace{1em}1999 J.Math.Phys. {\bf40} 2201\\
Znojil M\hspace{1em}1999 Phys.Lett.A {\bf259} 220\\
Znojil M\hspace{1em}2000 J.Phys.A: Math.Gen {\bf33} 4561\\
Ramirez A and Mielnit B\hspace{1em}2003 Rev.Fis.Mex. 49S2 130\\
Bender C M, Brody D C and Jones H F\hspace{1em}2003 Am.J.Phys.
{\bf71} 1095 \item Bagchi B
and Roychoudhury R\hspace{1em}2000 J.Phys.A: Math.Gen. {\bf33} L1\\
Bagchi B and Quesne C\hspace{1em}2000 Phys.Lett.A {\bf273} 285\\
Bagchi B and Quesne C\hspace{1em}2002 A{\bf300} 18\\
Bagchi B, Cannata F and Quesene C
\hspace{1em}2000 Phys.Lett.A {\bf209} 79\\
Znojil M, Cannata F, Bagchi B and Roychoudhury R\hspace{1em}2000 Phys.Lett.B {\bf483} 284\\
Bagchi B, Quesne C and Znojil M\hspace{1em}2001 Mod.Phys.Lett.A
{\bf16} 2047 \item Scholtz F G, Geyer H B and Hahne F J
W\hspace{1em}1992 Ann. Phys.(NY) {\bf213} 74 \item Dorey P,
Dunning C and  Tateo R\hspace{1em}2001 J.Phys.A: Math.Gen. {\bf34}
5679 \item Levai G and Znojil M\hspace{1em}2000 J.Phys.A:
Math.Gen. {\bf33} 7165 \item Kaushal R S\hspace{1em}2001 J.Phys.A:
Math.Gen. {\bf34} L709 \item Ahmed Z\hspace{1em}2001 Phys.Lett.A
{\bf282} 343 \item Andrianov A A, Ioffe M V, Cannata F and
Dedonder J P\hspace{1em}1999 Int.J.Mod.Phys.A {\bf14} 2675 \item
Kretschmer R and Szymanowski L\hspace{1em}2004 Phys.Lett.A
{\bf325} 112 \item Caliceti E, Graffi S and Maioli
M\hspace{1em}1980 Commun.Math.Phys. {\bf75} 51 \item Buslaev V and
Grecchi V\hspace{1em}1993 J.Phys.A: Math.Gen. {\bf26} 5541 \item
Mostafazadeh A\hspace{1em}- $\mathcal{PT}$ -symmetric cubic
anharmonic oscillator as a physical model, preprint
quant-ph/0411137 \item Bender C M, Brody D C and Jones H
F\hspace{1em}2002 Phys.Rev.Lett. {\bf89} 270401\item Bender C M,
Meisinger P N and Wang Q\hspace{1em}2003 J.Phys.A: Math.Gen.
{\bf36} 1973 \item Bagchi B, Banerjee A, Caliceti E, Cannata F,
Geyer H B, Quesne C and Znojil M\hspace{1em}- $\mathcal{CPT}$
-conserving Hamiltonians and their nonlinear supersymmetrization
using differential charge-operators $\mathcal{C}$ ,preprint
hep-th/0412211 \item Bender C M and Dunne G\hspace{1em}1989
Phys.Rev.D {\bf40} 2739 \item Mostafazadeh A\hspace{1em}2002
J.Math.Phys. {\bf43} 205\\Mostafazadeh A\hspace{1em}2002
J.Math.Phys. {\bf43} 2814\\Mostafazadeh A\hspace{1em}2002
J.Math.Phys. {\bf43} 3943 \item Mostafazadeh A,Batal
A\hspace{1em}2004 J.Phys.A: Math.Gen.{\bf37} 11645 \item
Mostafazadeh A - preprint quant-ph/0310164
\end{enumerate}

\end{document}